\documentclass[iop]{emulateapj}

\newcommand{\kms}{\textrm{km~s$^{-1}$}}

\newcommand{\ergs}{\textrm{erg~s$^{-1}$}}
\newcommand{\lsolar}{L$_{\odot}$}
\newcommand{\msolar}{M$_{\odot}$}
\newcommand{\ml}{M$_{\odot}$ yr$^{-1}$}
\newcommand{\mdot}{\dot{M}}

\usepackage{subfigure}
\usepackage{multirow}
\usepackage[dvips]{color}
\definecolor{Mygrey}{gray}{0.6}

\begin{document}
%\title{Evidence for Circumstellar Shells in Type Ia Supernovae}
\title{The Late-Time Rebrightening of Type Ia SN 2005gj in the Mid-Infrared}
%\title{The Rebrightening of Type Ia SN 2005gj in the Mid-Infrared as\\ Evidence of Multiple Circumstellar Shells}
%\titel{Late-Time ($>$300 day) Mid-Infrared Emission from Warm Dust in Type Ia Supernovae}
\shorttitle{SNe Type Ia Mid-IR Emission}
\author{Ori D. Fox\altaffilmark{1,2} \& Alexei V. Filippenko\altaffilmark{1}}
\altaffiltext{1}{Department of Astronomy, University of California, Berkeley}
\altaffiltext{2}{email: ofox@berkeley.edu .}

\begin{abstract}

A growing number of observations reveal a subset of Type Ia supernovae undergoing circumstellar interaction (SNe Ia-CSM).  We present unpublished archival $Spitzer~Space~Telescope$~data on SNe Ia-CSM 2002ic and 2005gj obtained $> 1300$ and 500 days post-discovery, respectively.  Both SNe show evidence for late-time mid-infrared (mid-IR) emission from warm dust.  The dust parameters are most consistent with a pre-existing dust shell that lies beyond the forward-shock radius, most likely radiatively heated by optical and X-ray emission continuously generated by late-time CSM interaction.  In the case of SN 2005gj, the mid-IR luminosity more than doubles after 1 year post-discovery.  While we are not aware of any late-time optical-wavelength observations at these epochs, we attribute this rebrightening to renewed shock interaction with a dense circumstellar shell.\\
\end{abstract}


\clearpage

\section{Introduction}
\label{sec:intro}

The ability to standardize Type Ia supernova (SN~Ia) light curves yields precise cosmological distance indicators \citep[e.g.,][]{phillips93}.  Despite these empirical relationships, questions remain about the underlying physics and progenitor systems.  The SN itself is generally accepted to be the thermonuclear explosion of a C/O white dwarf (WD), but the nature of the companion star remains ambiguous.  Evidence now exists for both single-degenerate and double-degenerate channels \citep[e.g.,][]{patat07,blondin09,simon09,nugent11,ganeshalingam11,brown12,foley12a,foley12b,bloom12,silverman13a}.

Recent studies reveal a subsample of SNe~Ia that exhibit signs of significant interaction between the forward shock and a dense circumstellar medium (hereafter SNe~Ia-CSM; \citealt{silverman13b}, and references within).  The spectra show many similarities to those of SNe~IIn, defined by their strong and relatively narrow hydrogen emission lines generated from a slow, dense, pre-existing CSM  (\citealt{schlegel90}; see \citealt{filippenko97} for a review).  The SNe~Ia-CSM also have slightly larger peak luminosities than typical SNe~Ia, in the range $-21.3 \leq M_R \leq -19$~mag.  Detailed observations of the nearby SN~Ia-CSM PTF11kx confirm a probable single-degenerate channel for at least this particular event \citep{dilday12,silverman13a}, although mass loss from a violent prompt merger after a common envelope (i.e., a core-degenerate scenario) has also been proposed \citep{soker13}.

Two of the most well-studied SNe~Ia-CSM are SNe 2002ic and 2005gj \citep{hamuy03,deng04,kotak04,wang04,wood-vasey04,aldering06,prieto07}.  The data suggest substantial mass loss (of order $>10^{-4}$~\ml) from the companion star and significant amounts of warm dust emitting in the near-infrared (near-IR).  These characteristics are also similar to those of SNe~IIn.  Owing to their dense CSM, SNe~IIn exhibit late-time ($> 100$ day) IR emission from warm dust that is continuously heated by visible and X-ray radiation generated by ongoing CSM interaction \citep[e.g.,][]{gerardy02, fox11}.  The CSM geometry derived from these dust shells reveals important clues about the progenitor mass-loss history.

In this {\it Letter} we present unpublished archival $Spitzer~Space~Telescope$~data on SNe~2002ic and 2005gj obtained $> 1300$ and 500 days post-discovery, respectively.  Section \ref{sec:2} lists the details of the observations; $Spitzer$ photometry constrains the dust mass and temperature, and thus the luminosity. We explore the origin and heating mechanism of the dust in \S \ref{sec:3}. Section \ref{sec:4} presents our conclusions.
\section{Observations}
\label{sec:2}

%% Table of Spitzer Observations
\begin{deluxetable*}{ l c c c c c c c c c c}
%\rotate
\tablewidth{0pt}
\tabletypesize{\footnotesize}
\tablecaption{$Spitzer$~Observations\tablenotemark{1}\label{tab1}}
\tablecolumns{9}
\tablehead{
\colhead{SN} & \colhead{JD} & \colhead{Epoch} & \colhead{PID} & \colhead{AOR} & \colhead{Distance} & \colhead{$t_{\rm int}$} & \colhead{3.6~\micron\tablenotemark{3}} & \colhead{4.5~\micron\tablenotemark{3}} & \colhead{5.8~\micron\tablenotemark{3}} & \colhead{8.0~\micron\tablenotemark{3}}\\
\colhead{}&\colhead{$-$2,450,000}&\colhead{(days)}&\colhead{}&\colhead{}&\colhead{(Mpc)}&\colhead{(s)} & \multicolumn{4}{c}{(10$^{17}$ erg s$^{-1}$ cm$^{-2}$ \AA$^{-1}$)}
}
\startdata
%Name & JD & Epoch & RA & DEC & Distance & tint
2002ic & 3386 & 795 & 3248 & 10550272 & 280 & 4000 & 0.150(0.015) & 0.136(0.012) & 0.092(0.008) & 0.052(0.004)\\
2002ic & 3770 & 1179 & 20256 & 14455040 &280 & 3600 & 0.060(0.009) & 0.063(0.008) & 0.052(0.006) & 0.040(0.004)\\
2002ic & 3961 & 1370 & 30292 & 17965824 & 280 & 2400 & 0.042(0.008) & 0.045(0.007) & 0.037(0.005) & 0.031(0.003)\\
2002ic & 4356 & 1765 & 40619 & 23107840 & 280 & 2400 & $<$0.03 & $<$0.03 & $<$0.03 & $<$0.03\\
2005gj & 3778 & 139 & 264 & 16868096 & 268 & 2400 & 0.130(0.014) & 0.092(0.010) & 0.036(0.005) & 0.017(0.003)\\
2005gj & 4004 & 365 & 30733 & 19308800 & 268 & 2400 & 0.183(0.016) & 0.110(0.011) & 0.049(0.006) & 0.024(0.003)\\
2005gj & 4149 & 510 & 30733 & 19309056 & 268 & 2400 & 0.276(0.020) & 0.199(0.014) & 0.113(0.009) & 0.049(0.004)
\enddata
\tablenotetext{1}{Upper limits for nondetections were derived by the point-source sensitivity in Table 2.10 of the IRAC Instrument Handbook, version 2.}
\tablenotetext{2}{All distances are derived from the host-galaxy redshift assuming $H_0$ = 72~km s$^{-1}$~Mpc$^{-1}$.}
\tablenotetext{3}{1$\sigma$~uncertainties are given in parentheses.}
\end{deluxetable*}

\subsection{Warm $Spitzer$/IRAC Photometry}
\label{sec:irac}

\begin{figure}
\begin{center}
\epsscale{0.9}
\subfigure[SN2002ic]{\label{f1a} \plotone{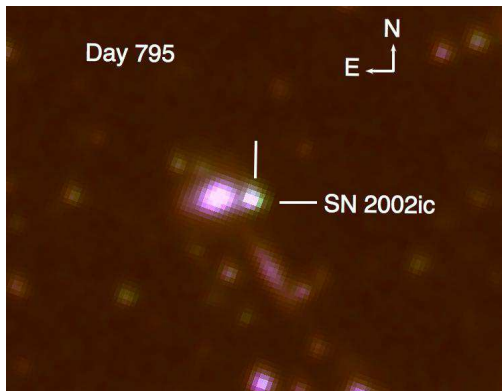}}\\
\subfigure[SN2005gj]{\label{f1b} \plotone{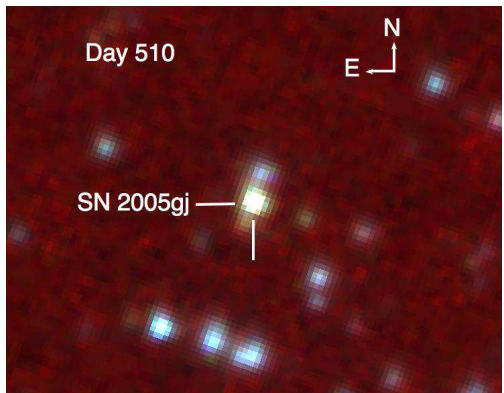}}
\caption{Combined false-color 3.6, 4.5, 5.8 \micron\ images of SNe 2002ic and 2005gj at late times ($\sim$1\arcsec$\times$1\arcsec).  Epochs refer to days post-discovery.
}
\label{f1}
\end{center}
\end{figure}

A search in the {\it Spitzer} Heritage Archive (SHA)\footnote{SHA can be accessed from http://sha.ipac.caltech.edu/applications/Spitzer/SHA/ .} revealed unpublished observations of SNe 2002ic and 2005gj, summarized in Table \ref{tab1}.  The SHA provide access to the Post Basic Calibrated Data ({\tt pbcd}), which are already fully coadded and calibrated.  Figure \ref{f1} shows false-color images of the combined 3.6, 4.5, and 5.8 \micron\ channel images at a single epoch.  The SN host galaxies tend to be bright and exhibit background-flux variations on rapid spatial scales.  Template subtraction can reduce photometric confusion from the underlying galaxy, but no pre-SN {\it Spitzer}~templates exist for these galaxies.  We therefore performed aperture photometry with {\tt DAOPHOT}/{\tt APPHOT} in {\tt IRAF}.\footnote{IRAF: the Image Reduction and Analysis Facility is distributed by the National Optical Astronomy Observatory, which is operated by the Association of Universities for Research in Astronomy (AURA) under cooperative agreement with the NSF.}  SN 2005gj falls $< 1$\arcsec\ from the galactic nucleus \citep{aldering06, prieto07}.  To minimize contributions from the underlying galaxy in both cases, a 2-pixel radius was chosen and aperture corrections were applied, although removal of nuclear contributions is admittedly difficult.

%% Spitzer Images
%\vspace{-20pt}

%% Spitzer Dust Fits
\begin{figure}
\begin{center}
\epsscale{1.}
\subfigure[]{\label{f2a} \plotone{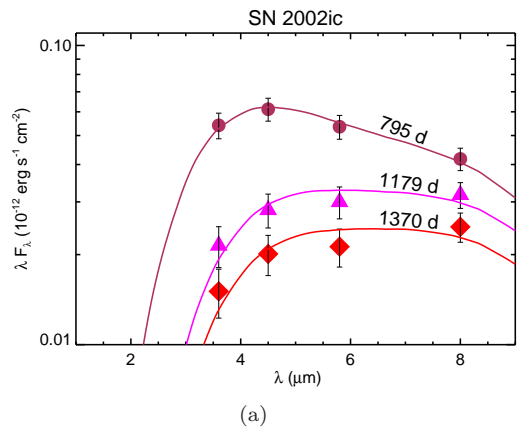}}\\
\subfigure[]{\label{f2b} \plotone{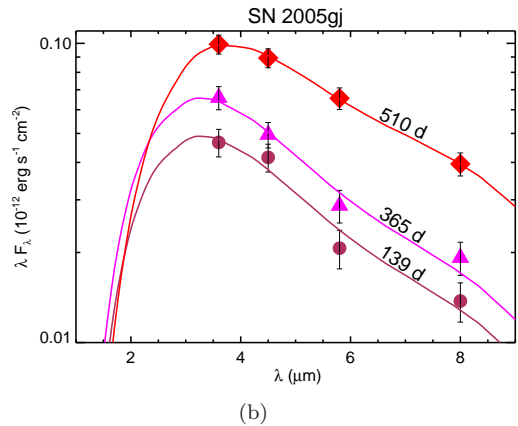}}
\caption{Photometry of SNe 2002ic and 2005gj in $Spitzer$/IRAC Channels 1 (3.6~\micron), 2 (4.5 \micron), 3 (5.8 \micron), and 4 (8.0 \micron).  Overplotted are the resulting best fits of Equation \ref{eqn:flux2}.
}
\label{f2}
\vspace{-15pt}
\end{center}
\end{figure}

\subsection{Dust Temperature and Mass}
\label{sec:dust}

Assuming thermal emission to be the dominant flux component, the mid-IR photometry probes the peak of the blackbody produced by warm grains.  The flux can be fit as a function of the dust temperature, $T_{\rm d}$, and mass, $M_{\rm d}$,
\begin{equation}
\label{eqn:flux2}
F_\nu = \frac{M_{\rm d} B_\nu(T_{\rm d}) \kappa_\nu(a)}{d^2},
\end{equation}
assuming optically thin dust with particle radius $a$, at a distance $d$ from the observer, thermally emitting at a single equilibrium temperature \citep[e.g.,][]{hildebrand83}, where $B_\nu(T_{\rm d})$~is the Planck blackbody function and $\kappa_\nu(a)$ is the dust absorption coefficient.  

For simplicity, we assume a simple dust population of a single size composed entirely of amorphous carbon (AC).  Figure \ref{f2} shows the best fit of Equation \ref{eqn:flux2} obtained with the IDL {\tt MPFIT} function \citep{markwardt09}, which minimizes the value of $\chi^2$ by varying $M_{\rm d}$ and $T_{\rm d}$.  With only four photometry points at each epoch, we limit our fits to a single component (see Figure \ref{f2}).  Table \ref{tab2} lists the best-fit parameters for AC grains of size $a = 0.1$ \micron.

\begin{deluxetable}{ l c c c c }
\tablewidth{0pt}
%\rotate
\tabletypesize{\footnotesize}
\tablecaption{IR~Fitting Parameters ($a = 0.1$ \micron~amorphous carbon) \label{tab2}}
\label{tab_phot}
\tablecolumns{5}
\tablehead{
\colhead{SN} & \colhead{Epoch (days)} & \colhead{$M_{\rm d}$ (\msolar)} & \colhead{$T_{\rm d}$} (K) & \colhead{$L_{\rm d}$ (\lsolar)}
}
\startdata
2002ic &  795  & 0.012  &  590    &  1.64$\times10^{8}$\\
2002ic &  1179 & 0.016  &  495     &  8.70$\times10^{7}$\\
2002ic &  1370 & 0.014  &  481     &  6.34$\times10^{7}$\\
2005gj &  139 & 0.001  &  845    &  1.06$\times10^{8}$\\
2005gj &  365 & 0.002  &  847    &  1.55$\times10^{8}$\\
2005gj &  510 & 0.006  &  725    &  2.34$\times10^{8}$
\enddata
\end{deluxetable}

\section{Analysis and Discussion}
\label{sec:3}

\begin{figure}[t]
%\rotate
\begin{center}
\plotone{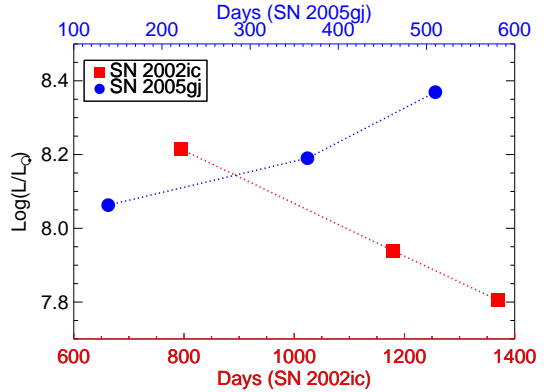}
\caption{The late-time mid-IR luminosity evolution of SNe 2002ic and 2005gj.  The luminosity is derived by summing over the modified blackbody fits given in Equation \ref{eqn:flux2} and shown in Figure \ref{f2}.  SN 2005gj exhibits an unanticipated rise more than a year post-discovery, suggesting renewed shock interaction between the forward shock and a pre-existing circumstellar shell.
}
\label{f3}
\end{center}
\end{figure}

Figure \ref{f3} plots the corresponding IR luminosity evolution for each object.  SN 2002ic remains bright ($L_{\rm d} > 10^{8}$~\lsolar) for more than 2~yr post-discovery, but continues to fade throughout the observations.  By contrast, SN 2005gj brightens after 1 year post-discovery.  While we only have mid-IR photometry available at these epochs, we explore the constraints these data provide on the circumstellar environment.

\subsection{Possible Origins and Heating Mechanisms}
\label{sec:origin}

The source of the mid-IR emission is warm dust, but the origin and heating mechanism of the dust are less clear.  The dust may be either newly formed or pre-existing, and either shock or radiatively heated; see \citet{fox10} for a full discussion.  To discriminate between possible scenarios, we assume a spherically symmetric, optically thin dust shell and calculate the blackbody radius, $r_{\rm bb} = [L_d/(4 \pi \sigma T_d^4)]^{1/2}$,
which sets a {\it minimum} shell size.  A luminosity $L_d \approx 10^{8}$~\lsolar\ and dust temperature $T_{\rm d} \approx 500$~K yield a blackbody radius of $r_{\rm bb} \approx 10^{17}$~cm.  This radius is larger than that of a forward shock moving at $v_{\rm s} \approx$ 10,000 \kms~on day 800.  This point seemingly rules out, particularly for the younger SN 2005gj, the possibility that the majority of the observed dust condensed in either the more slowly moving ejecta or the cold, dense shell that can form behind the forward shock. [We note, however, that \citet{silverman13b}~observed evidence of some newly formed dust in SN 2005gj via increasing absorption in the red H$\alpha$~wing.  While they do not estimate the dust mass, the $Spitzer$~observations suggest that this dust does not contribute significantly to the overall mid-IR flux.]

The alternative to newly formed dust is a pre-existing shell.  Again, the fact that the blackbody radius is beyond the forward-shock radius rules out the likelihood of shock heating.  Furthermore, an IR light-echo scenario \citep[e.g.,][]{dwek83}, in which the dust shell is heated by the peak SN luminosity, is unfeasible.  The implied shell radii ($r_{\rm echo} = t_{\rm plateau}/2{\rm c}$) would require peak luminosities of nearly $10^{11}$~\lsolar\ to heat the dust to the observed temperatures.

%%Light Curve

In fact, the observed dust-shell parameters (i.e., radius, temperature, mass) are comparable to those seen in SNe~IIn \citep[e.g.,][]{fox11, fox13}.  In these cases, ongoing interaction between the forward shock and dense CSM produce optical, UV, and X-ray emission that radiatively heat a pre-existing dust shell.  Assuming an optically thin dust shell, the observed dust temperature ($T_{\rm d}$) and shell radius ($r_{\rm d}$) require a combined optical, ultraviolet, and/or X-ray flux given by
\begin{equation}
\label{eqn:lbol}
L_{\rm opt/UV/X}  = \frac{64}{3} \rho a r_{\rm d}^2 \sigma T_{\rm SN}^4 \frac{\int{B_\nu (T_{\rm d}) \kappa(\nu) d\nu}}{\int{B_\nu(T_{\rm SN}) Q_{\rm abs}(\nu) d\nu}} 
\end{equation}
for a dust bulk (volume) density $\rho$ and an effective SN blackbody temperature $T_{\rm SN}$, where $B_\nu$~is the Planck blackbody function, $Q_{\rm abs}$ is the dust absorption efficiency, and $\kappa(\nu)$ is the dust absorption coefficient.  Figure 4 in \citet{fox13} shows that the blackbody radii of SNe 2002ic and 2005gj [$r_{\rm bb} \approx$ (0.5--1.0) $\times 10^{17}$~cm] at temperatures $T_{\rm d} \approx 500$--750~K require optical and/or X-ray luminosities in the range $10^8 \lesssim L_{\rm opt/UV/X} \lesssim 10^9$~\lsolar.  While we do not have optical or X-ray observations at these epochs, the most recent optical observations of SNe 2002ic and 2005gj are consistent, with measured luminosities of $10^{9.1}$~and $10^{9.4}$~\lsolar\ on days 250 and 149, respectively \citep{deng04,prieto07}.

\subsection{Evidence for Shells}
\label{sec:massloss}

Mid-IR wavelengths probe the characteristics of the CSM at the dust-shell radius.  Assuming a dust-to-gas mass ratio expected in the H-rich envelope of a massive star, $Z_{\rm d} = M_{\rm d}/M_{\rm g} \approx 0.01$, the dust-shell mass can be tied to the progenitor's total mass-loss rate, 
\begin{eqnarray}
\label{eqn:ml}
\mdot_{\rm outer} & = & \frac{M_{\rm d}}{Z_{\rm d} \Delta r} v_{\rm w} \nonumber \\ 
& = & \frac{3}{4} \Big(\frac{M_{\rm d}}{\rm M_{\odot}}\Big) \Big(\frac{v_{\rm w}}{120~\rm km~s^{-1}}\Big) \nonumber\\
&\times& \Big(\frac{5 \times 10^{16}~\rm cm}{r}\Big) \Big(\frac{r}{\Delta r}\Big) {\rm M}_{\odot}~{\rm yr}^{-1},
\end{eqnarray}
for a progenitor wind speed $v_{\rm w}$.  The relatively narrow lines observed in SNe~Ia-CSM originate in the slow pre-shocked CSM and can be used to approximate the progenitor wind speed.  The precursor wind velocities for SNe 2002ic and 2005gj are $v_{\rm w} =$ 100 and 60 \kms, respectively \citep{kotak04,aldering06}.  Assuming a thin shell, ${\Delta r}/r = 1/10$, wind speed $v_{\rm w} =$ 60 \kms, and radius $r_{\rm bb} = 10^{17}$~cm, the approximate mass-loss rate to produce the observed dust shell is $\mdot_{\rm outer} \approx 10^{-2}$~\ml.  A smaller dust-shell radius would require an even larger mass-loss rate.

The optical and/or X-ray emission generated by CSM interaction traces the mass loss at the inner radii.  Assuming a density $\propto r^{-2}$~wind profile, the rate can be written as a function of the optical/X-ray luminosity, progenitor wind speed, and shock velocity \citep[e.g.,][]{chugai94, smith09ip}:
\begin{eqnarray}
\label{eqn:inner}
\mdot_{\rm inner} & = & \frac{2 v_w}{\epsilon v_{s}^3}L_{\rm opt/UV/X}, \nonumber \\
 & = & 2.1 \times 10^{-4} \Big(\frac{L_{\rm opt/UV/X}}{3~\times 10^{41}~ \ergs}\Big) \nonumber\\
  &\times&\Big(\frac{\epsilon}{0.5}\Big)^{-1} \Big(\frac{v_w}{120~\kms}\Big) \nonumber\\
  &\times& \Big( \frac{v_s}{10^4~\kms}\Big)^{-3} {\rm M_{\odot}~yr^{-1}},
\end{eqnarray}
where $\epsilon < 1$~is the efficiency of converting shock kinetic energy into visual light.  We assume a value $\epsilon \approx 0.5$, although the conversion efficiency can vary with wind density and shock speed.  Again, we do not have late-time optical or X-ray observations, but we do have theoretical estimates from Equation \ref{eqn:lbol} in \S \ref{sec:origin}.  An optical luminosity $L_{\rm opt/UV/X} \approx 10^{8.5}$~\lsolar, wind speed $v_{\rm w} =$ 120 \kms, shock velocity $v_{\rm s} =$ 10,000 \kms~\citep{deng04}, and conversion efficiency $\epsilon=0.1$~correspond to a mass-loss rate $\mdot_{\rm inner} \approx 10^{-3}$~\ml.  

While the variables used above are only order-of-magnitude approximations, the derived rate for  $\mdot_{\rm inner}$ reveals two things about the circumstellar medium.  (1)  The difference between the mass-loss rates ($\mdot_{\rm inner} < \mdot_{\rm outer}$) suggests that the dust shells were formed during a period of increased, nonsteady mass loss.  (2) Compared to mass-loss rates derived from optical data at earlier epochs, our estimate of $\mdot_{\rm inner} \approx 10^{-3}$~\ml~is consistent with SN 2002ic on day 250 (for $\epsilon \approx 0.1$; \citealt{kotak04}), but at least an order of magnitude larger than that measured for SN 2005gj on day 74 \citep{prieto07}.  The increased circumstellar density derived for SN 2005gj at this late time suggests the presence of another shell of material.

The decline of SN 2002ic occurs at $>$800 days, corresponding to the time at which the forward shock ($v_{\rm s} =$ 10,000 \kms) would reach the blackbody radius ($r_{\rm bb} \approx 10^{17}$~cm).  The declining light curve may be attributed to the forward shock overtaking and destroying the dust and/or a decreasing amount of CSM interaction accompanied by a declining radiative heating source.

Alternatively, the rebrightening of the pre-existing dust in SN 2005gj is likely due to radiative heating by renewed shock interaction with this dense circumstellar shell.  From Equation \ref{eqn:inner}, an order-of-magnitude increase in mass loss results in an increase in the optical luminosity by a factor of 3--4, assuming the shock velocity decreases to 0.8 of its former value.  For a constant blackbody radius of $r_{\rm bb} \approx 5\times10^{16}$ cm, Equation \ref{eqn:lbol} and Figure 10 in \citet{fox10} show that an increase in the optical luminosity from $10^8$~to $10^{8.4}$~\lsolar\ results in a dust temperature increase from $\sim 600$ to 750~K.  Since $L_{\rm d} \propto T_{\rm d}^4$, this change in temperature results in a luminosity increase of a factor of 2.4. 

While we observe a luminosity increase of this magnitude (see Table \ref{tab2}), the inferred dust temperature actually {\it decreases}~by $\sim 120$~K.  The implication is that the blackbody radius must also increase from $5 \times 10^{16}$~to $10^{17}$~cm, which would explain a lower dust temperature along with a higher dust mass and luminosity.  The problem with invoking this scenario is that the only way to increase the blackbody radius (assuming a spherically symmetric shell of dust) would be for the increased optical luminosity to vaporize all dust out to the new blackbody radius.  For a dust vaporization temperature of $T_{\rm evap} \approx 2000$~K, Equation \ref{eqn:lbol} and Figure 8 of \citet{fox10} show that to vaporize dust out to $10^{17}$~cm requires a luminosity of $>10^{10}$~\lsolar, which is not likely.  The more probable explanation is that with only four data points, the slopes of the curves in Figure \ref{f2} and, therefore the temperatures, are biased by contamination from the underlying galaxy nucleus ($< 1$\arcsec\ away and a 0.6\arcsec~pixel scale).  The total dust luminosity, which is less sensitive to the slope, is most consistent with radiative heating by renewed shock interaction.

\section{Summary}
\label{sec:4}

This {\it Letter} presents unpublished archival $Spitzer$~data on SNe Ia-CSM 2002ic and 2005gj obtained $> 1300$ and 500 days post-discovery.  The mid-IR data show evidence of emission from warm dust.  While we do not have simultaneous observations at shorter wavelengths, the warm-dust parameters are most constant with a pre-existing dust shell heated by a combination of optical, UV, and X-ray emission continuously generated by ongoing CSM interaction.  The degree of CSM interaction dictates the dust temperature and, thereby, the luminosity.  In the case of SN 2005gj, the mid-IR luminosity nearly doubles more than 1 year post-discovery, suggesting an increasing amount of CSM interaction.  We attribute this renewed shock interaction to a dense circumstellar shell produced during a period of increased mass loss by the progenitor companion.  While progenitor mass loss suggests a single-degenerate channel, such rates have also been derived for both the core-degenerate and double-degenerate models \citep{livio03, ilkov12, ilkov13,soker13,shen13}.  Future multi-wavelength observations of SNe Ia-CSM will be necessary to better trace the complete mass-loss history and constrain the nature of the companion star.\\

\vspace{0 mm}

This work is based on archival data obtained with the {\it Spitzer Space Telescope}, which is operated by the Jet Propulsion Laboratory, California Institute of Technology, under a contract with NASA. Support for this work was provided by NASA through an award issued by JPL/Caltech (P90031). We also acknowledge generous financial assistance from the Christopher R. Redlich Fund, the TABASGO Foundation, and NSF grant AST-1211916.

\bibliographystyle{apj}

%\bibliography{references}

\end{document}